\documentclass[12pt]{article}

\usepackage{amsmath,amssymb,epsfig}
\usepackage[active]{srcltx}
\usepackage[sorted-cites,alphabetic]{amsrefs}

\newcommand{\be}{\begin{equation}}
\newcommand{\ba}{\begin{eqnarray}}
\newcommand{\ea}{\end{eqnarray}}

\def\ca{{\cal A}}
\def\cb{{\cal B}}

\def\cd{{\cal D}}

\def\cf{{\cal F}}

\def\ch{{\cal H}}

\def\cu{{\cal U}}

\newtheorem{thm}{Theorem}[subsection]

\newtheorem{prop}[thm]{Proposition}
\newtheorem{lemma}[thm]{Lemma}

\newtheorem{definition}[thm]{Definition}
\newtheorem{proposition}[thm]{Proposition}

\newcommand{\bbC}{{\Bbb C}}

\newcommand{\bbR}{{\Bbb R}}

\newcommand{\cF}{{\cal F}}

\fontfamily{yfrak}

\begin{document}

\vskip 25mm

\begin{center}

{\Large\bfseries 
$ C^*$-algebras of Holonomy-Diffeomorphisms   \\[1ex]\& Quantum Gravity
 II
}

\vskip 4ex

Johannes \textsc{Aastrup}$\,^{a}$\footnote{email: \texttt{aastrup@math.uni-hannover.de}} \&
Jesper M\o ller \textsc{Grimstrup}\,$^{b}$\footnote{email: \texttt{jesper.grimstrup@gmail.com}}\\ 
\vskip 3ex  

$^{a}\,$\textit{Institut f\"ur Analysis, Leibniz Universit\"at Hannover, \\ Welfengarten 1, 
D-30167 Hannover, Germany.}
\\[3ex]
$^{b}\,$\textit{Wildersgade 49b, 1408 Copenhagen K, Denmark.}\\[3ex]

\end{center}

\vskip 3ex

\begin{abstract}
We introduce the holonomy-diffeomorphism algebra, a $C^*$-algebra generated by flows of vectorfields and the compactly supported smooth functions on a manifold. We show that the separable representations of the holonomy-diffeomorphism algebra are given by measurable connections, and that the unitary equivalence of the representations corresponds to measured gauge equivalence of the measurable connections. We compare the setup to Loop Quantum Gravity and show that the
generalized connections found there are not contained in the spectrum
of the holonomy-diffeomorphism algebra in dimensions higher than one.


This is the second paper of two, where the prequel \cite{AastrupGrimstrup1} gives an exposition of a framework of quantum gravity based on the holonomy-diffeomorphism algebra.

\end{abstract}

\newpage
\tableofcontents

\section{Introduction}

Most fundamental theories of physics have connections among their basic variables,  like the standard model of particle physics and the Ashtekar formulation of general relativity. It is therefore important, especially with respect to a quantization of these theories, to consider functions of connections, i.e. observables, in these theories. Probably the most well known example of such functions are the Wilson loops, i.e. traces of the holonomies of loops along closed paths; but also open paths have been considered, in particular when the observables have to act on fermions.

One problem we have encountered in our attempt  \cite{AastrupGrimstruprew} to merge quantum gravity with noncommutative geometry  is that variables like the Wilson loops, and related variables, tend to discretize the underlying spaces. Therefore in this paper we will commence the study of an algebra of "functions" of smeared objects in order to avoid this discretization. More concretely we will study a $C^*$-algebra generated by flows of vector fields on a manifold $M$ and the smooth compactly supported functions on $M$. Flows of vector fields constitutes a natural notion of families of paths, and when evaluated on a connection in the spin-bundle $S$, naturally gives an operator on the spinors, i.e. on $L^2(M,S)$, and not like a path just acting on one point in $M$. The holonomy-diffeomorphism algebra is defined as the $C^*$-algebra generated by the flows and the smooth function with norm given by the supremum over all the smooth connections.   In this setup, smooth connections are viewed as representations of the holonomy-diffeomorphism algebra.      

It is the holonomy-diffeomorphism algebra which is our candidate for an algebra of observables.

One test to see if a found algebra of observables is suitable is to look at the spectrum of the algebra, i.e. the space of irreducible representations modulo unitary equivalence.
The main result in this paper is that all non-degenrate separable representations of the holonomy-diffeomorphism are given by  so called measurable connections. These are objects, which are similar to the generalized connections encountered in Loop Quantum Gravity (LQG), see \cite{AshtekarLewandowski}, but which take the measure class of the Riemannian metrics into account instead of the measure class of the counting measure.  The measurable  connections of course contain the smooth connections.

This paper is the second of two papers. Where its prequel \cite{AastrupGrimstrup1} is concerned with an exposition of a mathematical framework of quantum gravity based on the holonomy-diffeomorphism algebra this paper is solely concerned with the mathematical analysis of this algebra. \\

The paper is organized as follows:  

In section 2 we define the holonomy-diffeomorphism algebra. 

In section 3 we define the flow algebra. This algebra is constructed as a quotient of the cross product of the group generated the flows of the vector fields and the compactly supported smooth functions on the manifold. The ideal in this cross product, which is divided out, is the relation of local reparametrization. In particular the representations defining the holonomy-diffeomorphism algebra also give representations of the flow algebra.
We show that separable non degenerate representations of this flow algebra are given by so called measurable connections, and show the unitary equivalence between these measurable connection is given by measurable gauge equivalence.  

In section 4 we compare our setup with the LQG setup. The generalized connections appearing in LQG also gives rise to representations of the flow algebra, however non-separable representations. We show that the generalized connections can be obtained as the representations of a disretized version of the flow algebra. 

In section 5 we study the the properties of the representations of the holonomy-diffeomorphism algebra given by smooth connections. In particular we show that a connection is irreducible if and only if the corresponding representation of the holonomy-diffeomorphism algebra is irreducible, and give some structure of the separable part of the spectrum. 
In the second part of section 5 we show that if the dimension of the manifold is bigger than 1 the representations defined in section 4 coming from generalized connections are not contained in the spectrum of the holonomy-diffeomorphism algebra. We do, however, not know if there are other non-separable representations contained in the spectrum of the holonomy-diffeomorphism algebra.  \\

\textbf{Acknowledgement:} We thank M. Bekka, U. Haagerup and R. Nest for help concerning the representation theory of $Gl_n(\bbR )$.

We thank Adam Rennie for  enlightning discussions and great hospitality during our during our stay at the ANU, Canberra.

The first author would like to thank A. Tak\'acs and Horst S. for constant support.

\section{The holonomy-diffeomorphism algebra}
Let $M$ be a connected manifold and $S$ a vector bundle over $M$. We assume that $S$ is equipped with a fibre wise metric. This metric ensures that we have a Hilbert space $L^2 (M , \Omega^{\frac12} \otimes S)$, where $\Omega^{\frac12}$ denotes the bundle of half densities on $M$. Given a diffeomorphism $\phi: M\to M$ this acts unitarily on  $L^2 (M , \Omega^{\frac12} )$ through
$$ \phi (\xi)(m)= \varphi^*(\xi (\varphi (m) )  , $$
where 
$$\phi^* :\Omega^{\frac12} (\varphi (m)) \to \Omega^{\frac12} (m)  $$
denotes the pullback.

Let $X$ be a vectorfield on $M$, which can be exponentiated, and let $\nabla$ be a connection in $S$.  Denote by $t\to \exp_t(X)$ the corresponding flow. Given $m\in M$ let $\gamma$ be the curve  
$$\gamma (t)=\exp_{1-t} (X) (\exp_1 (X)(m) )$$
running from $\exp_1 (X)(m)$ to $m$. We define the operator 
$$e^X_\nabla :L^2 (M , \Omega^{\frac12} \otimes S) \to L^2 (M , \Omega^{\frac12} \otimes S)$$
in the following way:

Let $\xi \in L^2 (M , \Omega^{\frac12} \otimes S)$ be locally over $(\exp_1(x)(m)) $ of the form $f \otimes \omega \otimes s $, where $f$ is a function, $\omega$ an element in $\Omega^{\frac12}$ and $s$ an element in $S$.
The value of $(e^X_\nabla)(\xi)$ in the point  $m$ is given as
$$(f((\exp_1(X))(m)))  (\exp_1^*(\omega)) \otimes (\hbox{Hol}(\gamma , \nabla) s), $$
where $\hbox{Hol}(\gamma , \nabla)$ denotes the holonomy of $\nabla$ along $\gamma$.
If the connection $\nabla$ is unitary with respect to the metric on $S$, then  $e^X_\nabla$ is a unitary operator. 

If we are given a system of unitary connections $\ca$ we define an operator valued function over $\ca$ via
$$\ca \ni \nabla \to e^X_\nabla    ,$$
and denote this by $e^X$. Denote by $\cf (\ca , \cb (L^2(M,\Omega^{\frac12}\otimes S)) )$ the bounded operator valued functions over $\ca$. This forms a $C^*$-algebra with the norm
$$\| \Psi \| =  \sup_{\nabla \in \ca} \{\|  \Psi (\nabla )\| \}, \quad \Psi \in  \cf (\ca , \cb (L^2(M,\Omega^{\frac12}\otimes S)) ) $$ 
  
For a function $f\in C^\infty_c (M)$ we get another operator valued function $fe^X$ on $\ca$.

\begin{definition}
Let 
$$C =   \hbox{span} \{ fe^X |f\in C^\infty_c(M), \ X \hbox{ exponentiable vectorfield }\}  . $$
The holonomy-diffeomorphism algebra $\mathbf{H D} (M,S,\ca)   $ is defined to be the $C^*$-subalgebra of  $\cf (\ca , \cb (L^2(M,\Omega^{\frac12}\otimes S)) )$ generated by $C$.

We will by $\ch \cd (M,S,\ca)   $ denote the  $*$-algebra generated by $C$.
\end{definition}

It is this algebra that will be the object of study in this paper. We will in particular be interested in the spectrum and the representations of the holonomy-diffeomorphism algebra.

\subsection{Formulation with a metric}

The construction above of the holonomy-diffeomorphism algebra shows that it is background independent.  In some situations it is however convenient to have a formulation with a metric, and we will therefore in this subsection explain the construction given a fixed background metric $g$. 

Given an exponentiable vectorfield $X$ and a unitary connection $\nabla$, the first attempt to define an operator associated to $X$ would be to define
$$(e^X_\nabla ( \xi))( m)= \hbox{Hol} (\gamma , \nabla ) \xi ( \exp_1(X)(m)) ,\quad \xi \in L^2(M,S,dg)     .$$ 
(We have kept the notation from the previous section). The problem is, that since the flow of $X$ might not preserve the measure $dg$, this is not a unitary operator, and even worse it might not define an operator at all, since
$\| e^X_\nabla ( \xi )\|^2=\infty $ can occur.

To fix this we define the operator as
$$(e^X_\nabla ( \xi))( m)= \frac{\sqrt[4]{|g|} (\exp_1(X) (m))}{\sqrt[4]{|g|} (m)} \hbox{Hol} (\gamma , \nabla ) \xi ( \exp_1(X)(m)) ,\quad \xi \in L^2(M,S,dg)     .$$ 
This renders $e^X_\nabla$ unitary. We then define $\mathbf{HD}(M,S,\ca)$ and $\mathcal{HD}(M,S,\ca)$ as we did in the previous section.

\section{An abstract algebra}
In order to study the spectrum and the representations of $H D (M,S,\ca)   $ we will introduce an algebra which is more abstract than $H D (M,S,\ca)   $ and which at the end carries the information of the representation theory of $H D (M,S,\ca)   $ plus some additional information.  

Let $X$ be a vector field on $M$ which can be exponentiated. We will denote the flow to time $t$ as $e^{tX}$. By $e^X$ we denote the flow from $0$ to $1$, i.e. the map $ M\times [0,1] \to M   $ given by 
$$ (m,t)\to e^{tX} (m) .   $$
Two such flows for two vector fields $X_1,X_2$   can be composed via 
$$t \to \left\{ \begin{array}{cl} e^{(2t X_1)} &, t\in [0,\frac12 ]\\
e^{ ( (2t -1)X_2)}&,t \in [\frac12 , 1]
\end{array}    \right.$$
This is of course usually not a flow.  

If two flows are the same modulo reparametrization we will identify the flows. 
Also we will identify $e^Xe^{-X}$ with the trivial flow $I$, and $e^X \circ  I$, $I \circ e^X$ with $e^X$. We denote by $\cF$ the group generated by these flows. 

The flow group $\cf$ acts on $M$ simply by considering for $e^X$ the diffeomorphism $e^{1X}$. Note that this action is not faithful, since we are considering the diffeomorphism part of the flow. 

From the action on $M$ we induce a left action on $C^\infty_c(M)$ via
$$ e^X(f)(m)=f(e^{-1X}(m)) .  $$

We form the cross product
$$  \cF  \ltimes C^\infty_c (M). $$

 This algebra consists of the linear span of formal products 
$$ f F  , \quad f\in C^\infty_c (M), F\in \cF , $$ 
with the multiplication relation
$$f_1F_1 f_2 F_2=f_1 F_1 (f_2) F_1 F_2   $$
and adjoint
$$ (f_1 F_1)^*= F_1^*\overline{f}_1=F_1^{-1}(\overline{f}_1) F_1^{-1} .$$

If we are given a vector bundle $S$ with a metric over $M$ and a unitary connection $\nabla$ we get a $*$-representation $\varphi_\nabla$ of  $  \cF  \ltimes C^\infty_c (M) $ on $L^2(M,\Omega^{\frac12} \otimes S)   $ via 
$$\varphi_\nabla (fe^X )=fe^X_\nabla  .$$
Therefore unitary connections gives to representations of  $  \cF  \ltimes C^\infty_c (M) $. On the other hand a representation of $  \cF  \ltimes C^\infty_c (M) $ will in general not have much to do with unitary connections; the reason being that if $e^{X_1}\not= e^{X_2}$, but however coincide on $\hbox{supp}f$, then  $e^{X_1} f \not= e^{X_2} f$ in $\cF  \ltimes C^\infty_c (M)$, but $e^{X_1}_\nabla f = e^{X_2}_\nabla f$ for all connections.
We thus need to add extra relations to the cross product. 

Let $A$ be a subset of $M$. Let $F_1,F_2 \in \cF$. We will consider the $F_1$ and $F_2$ restricted to $A$ as maps 
$$F_1,F_2 : A\times [0,1] \to M . $$ 
We will say that $F_1$ is locally over $A$ a reparametrization of $F_2$ if for each $a\in A$ exists a monotone growing piecewise smooth bijection 
$$\varphi_a :[0,1] \to [0,1] $$  
such that 
$$ F_1 (a,t)=F_2(a,\varphi_a (t)) . $$  
\begin{definition}
Let $I$ be the subset of $\cF  \ltimes C^\infty_c (M)$ given by
$$\{F_1f-F_2 f  | F_1 \hbox{ is a local reparametrization of } F_2\hbox{ over \textnormal{supp}}f    \}  $$
\end{definition}

\begin{lemma}
$I$ is a $*$-ideal in $  \cF  \ltimes C^\infty_c (M) $.
\end{lemma}

\textit{Proof:} Multiplying $I$ from the right with a function $g\in C^\infty_C(M)$ preserves $I$. Multiplying from the left we get
$$g  (F_1f-F_2 f) =F_1 F_1^{-1}(g)f -F_2 F_2^{-1}(g)f    .$$
Since $F_1$ is a local reparametrization of $F_2$ we have $F_2^{-1}(g)f =F_1^{-1}(g)f$, and thus $g  (F_1f-F_2 f) \in I$.

When $F_1$ is a local reparametrization of $F_2$, then so is $FF_1$ of $FF_2$. Hence multiplication from the right with $\cF$ preserves $I$.

Right multiplication yields
$$(F_1f-F_2 f)F=(F_1F -F_2F)F(f) ,       $$
and since $F_1F$ is a local reparametrization of $  F_2F$ over supp$F(f)$ we have $IF \subset I$. \hfill $\Box$

\begin{definition}
The flow algebra $\cF M$ is defined as
$$ \cF  \ltimes C^\infty_c (M) / I.  $$
\end{definition} 

Note that for  a unitary connection $\nabla$ we have $\varphi_\nabla (I)=0$, and hence it descends to a $*$-representation, also denoted $\varphi_\nabla$, of $\cF M$  on $L^2(M,\Omega^{\frac12}\otimes S)$. 

It follows that $C^\infty_c (M)$ embeds into $\cF M$ as $f\to f \cdot e^0$.

\subsection{Separable representations}

We will now study the separable $*$-representations of $\cF M$, i.e. $*$-homomorphisms from $\cF M$ to $\cb (\ch ) $, $\ch$ separable. 
 
 Therefore let $\varphi : \cF M \to \cb (\ch )$ be such a representation, and we will also assume that $\varphi$ is non-degenerate. In particular we get a $*$-representation, also denoted $\varphi$, of $C_c^\infty (M)$ on $\ch$.
 
\begin{lemma}
Let $\varphi : C^\infty_c (M) \to  \cb (\ch )$ be a $*$-representation. Then $\varphi$ has a unique extension to a $*$-representation 
 $$\tilde{\varphi} :C_0(M)\to   \cb ( \ch).$$
\end{lemma}

\textit{Proof:} We only need to show that $\varphi$ is continuous. We first extend $\varphi$ to a unital $*$-homomorphism 
$$\varphi^\sim :  C^\infty_c (M)^\sim  \to  \cb (\ch ) , $$
where $C^\infty_c (M)^\sim$ denotes $C^\infty_c (M) $ with a unit added. The norm of $\varphi^\sim (f)$ is given via the spectral radius. If $\lambda \notin f(M)$ then $\lambda-f$ has an inverse in  $C^\infty_c (M)^\sim$. This implies that $\lambda \notin \hbox{spec}(\varphi^\sim (f))$, and $\varphi$ is therefore continuous. \hfill $\Box$   
\\

We will also denote the extension by $\varphi$. The commutator $\varphi (C(M))' $ of $\varphi (C(M))$ in $\cb (\ch )$ is a type $I$ von Neumann algebra. This means that there exists an orthogonal family of projections $\{ P_n  \}_{n\in \{ 1, \ldots , \infty \} }$ in the center of $\varphi (C(M))' $ with $\sum P_n=\mathbf{1_\ch}$ such that $P_n \varphi (C(M))' $ is a type $I_n$.     

Let $F\in \cF$ and $A\in \varphi (C(M))' $. Then $\varphi (F)A \varphi (F^{-1}) \in \varphi (C(M))' $ since 
\begin{eqnarray*}
\lefteqn{\varphi (f) \varphi (F)A \varphi (F^{-1})} \\
&  =&   \varphi (F) \varphi (F(f))A \varphi (F^{-1}) =\varphi (F)  A \varphi (F(f)) \varphi (F^{-1})= \varphi (F)A \varphi (F^{-1}) \varphi (f) .  
\end{eqnarray*}

Conjugating with $\varphi (F)$ is therefore an automorphism of $\varphi (C(M))' $, and it follows that conjugating with $\varphi (F)$ preserves the type structure, and thus  $\varphi (F)$ commutes with each $P_n$. This means that studying the representations of $\cf M$ we can restrict to the case of $\varphi (C(M))' $ being of type $I_n$ for a fixed $n\in \{ 1, \ldots , \infty \}$. In fact it follows from what will come that all representations are of type $I_n$ for a fixed $n$. 

Since $\varphi (C(M))' $ is of type $I_n$, $\varphi (C(M))' $ is isomorphic to $L^\infty (X)\otimes \cb (\ch ) $, with dim$\ch=n$, and $X$ being some $\sigma$-finite measure space with measure $\mu$. This means that the representation $\varphi$ is given via a representation 
$$\psi: C^\infty_c (M)\to \cb (L^2 (X,\mu )) $$
with $\psi   (C^\infty_c (M) )\subset L^\infty (X)$ and $\varphi =\psi \otimes \mathbf{1}$. Since $\varphi$ is non-degenerate, we can assume that $X=M$,  $(\psi (f)\xi )(m)=f(m)\xi (m)$, and that $\mu$ is a regular Borel measure on $M$. Furthermore we can also assume that $(M,\mu)$ is finite. 

All together we have
\begin{proposition}
Let $\varphi: \cf M \to \cb (\ch )$ be separable non-degenerate representation. There exists a finite Borel measure $\mu$ on $M$ such that $\varphi $ is unitarily equivalent to a representation $\varphi_1$ on $L^2 (M,\mu )\otimes \ch_n$, dim$\ch=n$, $n=1,\ldots ,\infty$, with 
$$(\varphi_1 (f)\xi\otimes \eta)(m)=f(m) \xi (m) \eta, \quad f\in C^\infty_c(M).  $$
\end{proposition}
\vskip 1cm

Let $\varphi$ be a representation of the same form as $\varphi_1$ in the above proposition. We will identify $L^2 (M,\mu )\otimes \ch_n$ with $L^2 (M,\mu , \ch_n)$. We assume, that $\mu (M)=1$. We define the measure $\mu_F$ as
$$\mu_F(A)=\mu (F (A)) , $$
for all  measurable subsets $A$ of $M$.

Let $F$ be a flow and let $f\in C_0(M)$. We have
\begin{equation} \label{id}
(\varphi (F) \varphi (f) \xi )(m) = (F(f)) (m) (\varphi (F) \xi )(m)  .
\end{equation}
Let $\xi$ be a vector with $\| \xi (m)\|_{2,n}=1$, where $\| \cdot \|_{2,n}$ denotes the norm on $\ch_n$. We define
$$k_F(m)=\| \varphi (F) (\xi)(m)  \|^2_{2,n} .$$
This is independet of $\xi$ $\mu$-almost everywhere because
\begin{eqnarray} \label{formel}
\| f\|^2_2 &=& \langle \varphi (f) \xi ,\varphi (f)\xi \rangle =\langle \varphi (F) \varphi (f) \xi , \varphi (F) \varphi (f)\xi \rangle  \nonumber \\
&=& \langle \varphi (F(f)) \varphi (F)  \xi , \varphi (F(f)) \varphi (F)\xi \rangle  \nonumber  \\
&=  & \int_M |f(F^{-1}(m) |^2\| \varphi (F) \xi (m)\|_{2,n}^2  d\mu (m)  
\end{eqnarray}
for all $f\in C_0(M)$.





Since formula (\ref{formel}) holds for all bounded measurable functions we get
$$\mu_F (F^{-1}(A)) =\mu (A)=\int_M 1_A^2 d\mu =  \int_M |F(1_A)|^2 k_F d\mu = \int_M 1_{F^{-1}(A)} k_F d\mu , $$
thus $\mu_F=k_F \mu$, and therefore $\mu_F \ll  \mu $. The same holds for $F^{-1}$ and we get

\begin{lemma} \label{aekvi}
$\mu$ is equivalent to $\mu_F$.
\end{lemma}

With this lemma we can now prove

\begin{thm}
$\mu$ is equivalent to a measure induced by a Riemannian metric on $M$.
\end{thm}

\textit{Proof:} This follows from lemma \ref{aekvi} and the result that a quasi-ivariant Borel  measure on a locally compact group is equivalent to the Haar measure.   
\hfill $\Box$   \\ 

Because of (\ref{id}) we can consider $\varphi (F)$ as a measurable family $m\to \varphi (F)(m)$ in $U(n)$, the group of unitary operators on $\ch_n$, i.e. 
$$(\varphi (F)\xi ) (m) =\sqrt{k_F (m)} \varphi (F)(m) (\xi (F^{-1}(m)) .$$

This gives rise to the following
\begin{definition}
A measurable $U(n)$-connection, $n=1,\ldots , \infty$, is a map $\nabla$ from $\cF$ to the group of measurable maps from $M$ to $U(n)$ satisfying
\begin{enumerate}
\item $\nabla (1)= 1$.
\item $\nabla (F_1 \circ F_2)(m)=\nabla (F_1) (m) \circ \nabla (F_2)(F_1^{-1}(m))$
\item If $F_1$ and $F_2$ are the same up to local reparametrization over some set $U\subset M$, then 
$$ \nabla ( F_1)_U= \nabla (F_2)_U  . $$ 
\end{enumerate} 
\end{definition}

A measurable $U(n)$-connection $\nabla$ gives rise to a representation of $\cF M$ on $L^2(M,\Omega^{\frac12}\otimes \ch_n )$ via
$$  ( \varphi_\nabla (f)(\xi))(m)=f(m)\xi (m) $$
$$  \varphi_\nabla (F)(\xi)(m) = \big(   (F^{-1})^* (F^{-1}(m))  \otimes  \nabla (F)(m)) \big)\xi (F^{-1}(m))    $$

With the work done so far in this section we have 

\begin{thm} \label{flowrep}
Any non-degenerate separable representation of $\cF M$ is unitarily equivalent to a representation of the form $\varphi_\nabla$, where $\nabla$ is a measurable $U(n)$-connection.
\end{thm}

\subsection{Equivalence of representations}
\begin{definition}
Two representations $\varphi_1, \varphi_2$ of $\cf M$ on $\ch_1,\ch_2$ are called unitary equivalent if there exist a unitary $U:\ch_1 \to \ch_2$ with  
$$\varphi_1(a)=U^* \varphi_2 (a) U , \quad \hbox{ for all } a\in \cf M. $$  
\end{definition}

According to \ref{flowrep} any two non-degenrate separable representations are of the form $\varphi_{\nabla_1}, \varphi_{\nabla_2}$ and  $\ch_1=L^2(M,\Omega^{\frac12}\otimes \ch_{n_1} )$, $\ch_2=L^2(M,\Omega^{\frac12}\otimes \ch_{n_2} )$. 
If they are unitarily equivalent we have 
$$(\varphi_{\nabla_1} (f)\xi )(m)=f(m)\xi (m)=(\varphi_{\nabla_2} (f)\xi )(m), $$
and therefore  the unitary $U$ is given as a measurable map $m\to u(m)$, with $u(m):\ch_{n_1} \to \ch_{n_2}$ unitary. Consequently we have $n_1=n_2$. 

This leads to the following
\begin{definition}
A measurable $U(n)$-gauge transform is a measurable map
$$ M\ni m \to \cu (\ch_n )  .$$

Two measurable $U(n)$-connections $\nabla_1$, $\nabla_2$ are called gauge equivalent if there exists a measurable $U(n)$-gauge transform $m\to u(m)$ with   
$$ \nabla_1(F)(m)=u(m) \nabla_2(F)(u(F^{-1}(m)))^*\hbox{ for all }F\in \cF .    $$
\end{definition}

\begin{prop}
Two representation of the form $\varphi_{\nabla_1}$, $\varphi_{\nabla_2}$, $\nabla_1 , \nabla_2$ measurable $U(n)$ connections, are unitarily equivalent if and only if $\nabla_1 , \nabla_2$ are measurable gauge equivalent.
\end{prop}


\section{Comparision to  the {{LQG}} spectrum}
In this section we will compare the setup we have so far to that of LQG, and also give some   non-separable representation of $\cf M$. For simplicity we will only work with piecewise analytic flows and paths. When we talk about paths we will identify $l^{-1} \circ l$ with the trivial path starting and ending in the start point of $l$. We will also identify two path which are the same up to reparametrization. 

\begin{definition}
Let $G$ be a connected Lie-group. A generalized connection is an assignment $\nabla (l)\in G$ to each piecewise analytic edge $l$, such that 
$$\nabla (l_1 \circ l_2 )=\nabla (l_1) \nabla (l_2) .$$

\end{definition}

For details on generalized connections see \cite{AshtekarLewandowski1,MarolfMourao}, see also \cite{AastrupGrimstrup2}. 

Let us now further assume that we have a  representation of $G$ as a subgroup of $U(n)$. Note that we can in general not  use a generalized connection to define a representation of $\cf M$ on $L^2(M, \Omega^{\frac12} \otimes \ch_n )$ like we did for a smooth or measurable connection. The problem is, that $e^X_\nabla$ need not  be measurable. 

On the other hand, if we equip $M$ with the counting measure, we can use a generalized connection $\nabla$ to define a representation of $\cf M$
on $L^2(M, \ch_n)$. Here we however see, that a measurable connection does not define a represention on $L^2(M,\ch_n)$, since a measurable connection is only defined up to zero sets, and therefore not in single points.

\begin{definition}
A generalized unitary gauge transformation  is a map 
$$U:M\to U(n).$$
Two generalized connections $\nabla_1 $ and $\nabla_2$ are said to be unitarily gauge equivalent if for all paths $l$ we have
$$U(e(l)) \nabla_1(l)U^*(s(l))  =\nabla_2(l),  $$
where $e(l)$ denotes the end point of $l$ and $s(l)$ the start point.
\end{definition}

In order to see the generalized connections as related to the spectrum of an algebra similar the the flow algebra, we will define a discrete version of it. 

Let $C_d(M)$ be the algebra of functions on $M$ with finite support. We define $\cf_d M$ like $\cF M$ but with $C^\infty_c (M)$ replaced by $C_d (M)$. For a given point $m\in M$ we denote by $1_m$ the function with value $1$ in $m$ and zero elsewhere. This is a projection, and due to the relations defining $\cF_dM$ we have $F 1_m F^{-1}=1_{F(m)}  $.

Given a non-degenerate representation $\varphi :\cF_dM \to \cb (\ch ) $ we define $\ch_m=\varphi (1_m)\ch$. Since $\varphi (F)$ is a unitary operator, then, due to the conjugation relation with $1_m$ above, $\varphi (F) :\ch_m \to \ch_{F(m)}$ is a unitary operator. In particular we have 
$$\ch =\bigoplus_{m\in M}   \ch_m ,$$
and all the $\ch_m$'s have the same dimension, and we can therefore write the Hilbert space as
$$ \ch =\bigoplus_{m\in M}   \ch_n  =L^2(M,\ch_n),$$ 
where $M$ is equiped with the counting measure and $\ch_n$ is a Hilbert space of dimension $n$, $n$ being a cardinal number.

We thus have
\begin{proposition}
To every non-degerate representation $\varphi$ of $\cf_d M$, there exists a generalized connection $\nabla$, such that $\varphi$  is of the form 
$$\varphi : \cf_dM  \to \cb (L^2 (M,\ch_n)) ,$$
with
$$(\varphi (f) \xi ) (m)=f(m)\xi (m) , $$
and 
$$(\varphi (F) \xi ) ( F(m))=\nabla (F_m) \xi (m) ,  $$
where $F_m$ is the edge $F$ defines between $m$ and $F(m)$.

Two non-degenrate rapresentations $\varphi_1, \varphi_2$, associated to two generalized connections $\nabla_1,\nabla_2$, are equivalent if and only if they are generalized gauge equivalent. 
\end{proposition}

\section{The spectrum of the holonomy-diffeomorphism algebra}

We will in this section restrict attention to $\mathbf{H D} (M,S,\ca)$, where $S$ is the trivial two-dimensional bundle, and $\ca$ is the set of $SU(2)$-connections.  

\subsection{Properties of the representations}

We remind the reader of the following two 
\begin{definition}
A connection is called irreducible if in a given point $m$ the holonomy group in this point  acts irreducible on the fibre in $m$. 
\end{definition}

\begin{definition}
A $*$-representation $\varphi :A \to \cb $ of a $C^*$-algebra $A$ on a Hilbert space $\ch$ is called irreducible if $\varphi (A)$ acts irreducibly on $\ch$.
\end{definition}

There is the following well known charaterization of irreducible representations, see \cite{BratteliRobinson}

\begin{thm}
A representation $\varphi: A\to \cb (\ch)$ of a $C^*$-algebra on a Hilbert space is called irreducible if one of the following equivalent conditions are fullfilled:
\begin{enumerate}
\item $\varphi $ is irreducible.
\item The commutant $\varphi (A)'=\{ b\in \cb (\ch)| b\varphi (A)=\varphi (A)b \hbox{ for all }a\in A\}$ is equal to $\bbC 1_\ch$.
\item Every nonzero vector $\xi \in \ch$ is cyclic for $\varphi (A)$, or $\varphi (A)=0$ and $\ch =\bbC$. 
\end{enumerate}
\end{thm}

We can now connect the two notions of irreducibility
\begin{proposition}
A representation $\varphi_\nabla$ is irreducible if and only if $\nabla$ is irreducible.

When $\nabla$ is reducible, the representation $\varphi_\nabla$ splits into two irreducible representations, corresponding to $U(1)$ connections.
\end{proposition}

\textit{Proof:} Clearly the representations $\varphi_\nabla$ are not zero. 

Let us assume that $\nabla$ is irreducible. Since the holonomy group acts irreducibly in one point it acts irreducibly in all points, and for every two points in $M$ the holonomy paths between these two points acts irreducibly between the fibers in the points.  Since we have the flows as operators, and we can multiply these with compactly supported smooth functions, it follows that every nonzero vector is cyclic for $\varphi_\nabla (\mathbf{H D} (M,S,\ca))$, i.e.  $\varphi_\nabla $  acts irreducibly. 

On the other hand, if $\nabla$ is not irreducible, we can split the bundle $S$ into two line bundle, each being invariant under the action of the holonomy groupoid. Consequently $\varphi_\nabla$ splits into two irreducible representations, each corresponding to a $U(1)$ connection.
\hfill $\Box$   \\ 

This motivates the following
\begin{definition}
A measurable $U(n)$-connection $\nabla$ is called irreducible if the corresponding representation $\varphi_\nabla$ is irreducible.
\end{definition}

We remind the reader of the following

\begin{definition}
Let $A$ be a $C^*$-algebra. The spectrum of $A$, $\hbox{spec}( A)$, is defined as
$$ \hbox{spec}(A) =\{ \hbox{Irreducible representations} \} / \hbox{Unitary equivalence }   . $$ 
\end{definition}

We have
\begin{proposition} Put
$$  \mathcal{U}_1=\{ \hbox{Measurable }U(1)\hbox{-connections} \}   $$
and 
$$  \mathcal{U}_2=\{ \hbox{Irreducible measurable }U(2)\hbox{-connections} \}  . $$
The separable part of the $\hbox{spec} (\mathbf{H D} (M,S,\ca))$ is contained in 
$$ ( \mathcal{U}_1 \cup \mathcal{U}_2)/\{ \hbox{Measurable gauge equivalence }   \}   $$
\end{proposition}

\textit{Proof:} The only statement that remains to prove is that irreducible measurable $U(n)$-connections, $n\geq 3$, do not appear in the spectrum. This follows, since representations of rank below $2$ form a closed subset in the spectrum. \hfill $\Box$ \\

\subsection{The non-separable part of the spectrum of the holonomy-diffeomorphism algebra}

In general we do not know what the non-separable part of the spectrum looks like. We have, however, the following
\begin{proposition}
Let \text{dim}$(M)>1$. The representations $\psi_\nabla $, where $\nabla$ is a generalized connection  given by representing the flow-algebra on $L^2(M,\ch_n)$ with the counting measure,  are not contained in the spectrum of $\mathbf{H D} (M,S,\ca)$.
\end{proposition}

\textit{Proof:} We will show that $\psi_\nabla$ can not be bounded by  the representations of the form $\varphi_{\nabla_1}: \cf M \to \cb (L^2 (M,\Omega^{\frac12}\otimes S)) $, where $\nabla_1$ is a smooth connection. 

We choose an open subset $U$ of $M$ diffeomorphic to $\bbR^n$, $n\geq 2$.
 We consider a subgroup of the flow group, which acts on $U$ like $Gl_n^+(\bbR)$ on $\bbR^n$. 

The $Gl_n^+(\bbR)$-part of the representation $\psi_\nabla$ is given by representing $Gl_n^+(\bbR)$ on $L^2(\bbR^n, c)$, $c$ being the counting measure. The subspace $\bbC 1_0$ of $L^2(\bbR^n, c)$ is invariant under this representation, and therefore the trivial represenation of $Gl_n^+(\bbR)$ is contained in this representation.

The representation $\varphi_\nabla$ is equivalent to two copies of  a representation $\pi$ on $L^2(\bbR^n)$, where the $Gl_n^+(\bbR)$-part of the representation is  given by 
 $$ (\pi(g)(\xi)(x)=|\hbox{det}g|^{-\frac12} \xi ( xg^{-1} ), \quad g\in Gl_n^+(\bbR) .   $$

The $n$-fold tensor product of $\pi$ is given by
$$  (\pi^{\otimes n} (\xi))(x)=  |\hbox{det}g|^{-\frac{n}{2}} \xi (xg^{-1}), \quad x\in M_n(\bbR) $$
on $L^2(M_n(\bbR), \lambda )$. Since the Haar measure on $Gl_n(\bbR) $ 
is given by 
$$\mu (S) =\int_S |\hbox{det}x|^{-n} dx,      $$
where $S$ is a subset of $M_n(\bbR)$, the $n$-fold tensor product of $\pi$ is equivalent to the left regular representation of $Gl_n^+(\bbR)$ on $L^2(Gl_n(\bbR), \mu )$. In particular $\pi$ is bounded by the left regular representation. 

We will now show, that the trivial representation of $Gl_n^+(\bbR)$ is not weakly contained in the left regular representation, thereby concluding that $\psi_\nabla$ can not be bounded by  the representations of the form $\varphi_{\nabla_1}$, where $\nabla_1$ is the trivial connection on $\bbR^n$. 

Denote by $\lambda$ the left regular representation of  $ Gl_n^+ (\bbR )$ on $L^2 (Gl_n(\bbR ))$.  We want to show that the trivial representation is not weakly contained in this representation when we consider $Gl_n^+(\bbR )$ as discrete group. Since $\lambda \otimes \overline{\lambda} $ is equivalent to a multiple of $\lambda$ it follows that the trivial representation is weakly contained in $\lambda$ if and only if the trivial representation is weakly contained in   $\lambda \otimes \overline{\lambda} $. According to Theorem 5.1 in \cite{Bekka} this is equivalent to the trivial representation being weakly contained in  $\lambda \otimes \overline{\lambda} $ when $Gl_n^+(\bbR )$ is considered as a continuous group with the usual topology. This is then again equivalent to the trivial representation being weakly contained in $\lambda$   when $Gl_n^+(\bbR )$ is considered as a continuous group with the usual topology. Restricting to $\lambda$ to $Gl_n^+(\bbR )$ we get that the trivial representation is contained in the left regular representation of  $Gl_n^+(\bbR )$ as continuous group.   This is however in contradiction to the $Gl_n^+(\bbR )$ being non-amenable  for $n\geq 2$, see \cite{Greenleaf}. 

If $\nabla_2$ is an arbitrary  smooth $SU(2)$-connection we can proceed as follows: Since the trivial representation of $Gl_n^+(\bbR)$ is not weakly contained in the left regular representation, there exists, according to proposition G.4.2 in \cite{Bekka2},   positive numbers $a_1,\ldots a_k$ and elements $g_1,\ldots , g_k$ in the flow group with 
$$ \| \varphi_{\nabla_1}(a_1g_1+\ldots +a_kg_k)\| <\|\psi_\nabla (a_1g_1+\ldots +a_kg_k)\|  .  $$
However since $a_1,\ldots ,a_k$ are positive, we have 
$$ \| \varphi_{\nabla_2}(a_1g_1+\ldots +a_kg_k)\| \leq \| \varphi_{\nabla_1}(a_1g_1+\ldots +a_kg_k)\|  .$$
Hence $\psi_\nabla $ can not be bounded by  $\varphi_{\nabla_2}$.\hfill $\Box$

\end{document}